\documentclass{article}
\usepackage{jheppub}

\usepackage{amsfonts}

\def\det{{\rm det\,}}
\def\re{{\rm Re\,}}
\def\im{{\rm Im\,}}

\def\sign{{\rm sign\,}}

\def\bea{\begin{eqnarray}}
\def\eea{\end{eqnarray}}
\def\nn{\nonumber}

\def\lmatrix{\left(\begin{array}}
\def\rmatrix{\end{array}\right)}
\def\gsim{\mathrel{\rlap{\lower4pt\hbox{\hskip1pt$\sim$}}\raise1pt\hbox{$>$}}}
\def\lsim{\mathrel{\rlap{\lower4pt\hbox{\hskip1pt$\sim$}}\raise1pt\hbox{$<$}}}
\def\bi{\begin{itemize}}
\def\ei{\end{itemize}}
\def\msbar{\overline{\rm MS\kern-0.5pt}\kern0.5pt}
\def\rho{\varrho}

\title{New approach to lattice QCD at finite density; results for the critical end point on coarse lattices}

\author{Matteo Giordano, }
\author{Kornel Kapas, }
\author{Sandor D. Katz, }
\author{Daniel Nogradi, }
\author{Attila Pasztor}

\affiliation{ELTE Eotvos Lorand University, Department of Theoretical Physics, Pazmany Peter setany 1/a, Budapest 1117, Hungary}

\emailAdd{giordano@bodri.elte.hu}
\emailAdd{kapaskornel@caesar.elte.hu}
\emailAdd{katz@bodri.elte.hu}
\emailAdd{nogradi@bodri.elte.hu}
\emailAdd{apasztor@bodri.elte.hu}

\abstract{
All approaches currently used to study finite baryon density lattice QCD suffer from uncontrolled systematic uncertainties
in addition to the well-known sign problem.
We formulate and test an algorithm, sign reweighting, that works directly at finite $\mu =
\mu_B/3$ and
is yet free from any such uncontrolled systematics. With this algorithm the {\em only} problem is the sign problem itself.
This approach involves the
generation of configurations with the positive fermionic weight $|\re\det D(\mu)|$ where $D(\mu)$ is the Dirac matrix
and the signs ${\rm sign} ( \re \det D(\mu) ) = \pm 1$ are handled by a discrete reweighting.
Hence there are only
two sectors, $+1$ and $-1$ and as long as the average $\langle\pm 1\rangle \neq 0$ (with respect to the positive weight) 
this discrete reweighting by the signs carries no overlap
problem and the results are reliable. The approach is tested on $N_t = 4$ lattices with $2+1$ flavors 
and physical quark masses using the unimproved staggered discretization. By measuring the Fisher 
(sometimes also called Lee-Yang) zeros in the bare coupling
on spatial lattices $L/a = 8, 10, 12$ 
we conclude that the cross-over present at $\mu = 0$ becomes stronger at $\mu > 0$ and is consistent with a
true phase transition at around $\mu_B/T \sim 2.4$.
}

\begin{document}

\maketitle

\section{Introduction}
\label{introduction}

The numerical simulation of lattice QCD at finite baryon chemical potential is known to be hindered by the notorious
sign problem: the fermionic determinant is not real and hence importance sampling techniques do not apply. Ways around
the problem were nonetheless devised. These include Taylor expansion 
\cite{Allton:2002zi,Gavai:2003mf,Gavai:2004sd,Allton:2005gk,Gavai:2008zr,
Basak:2009uv,Borsanyi:2011sw,Borsanyi:2012cr,Bellwied:2015lba,
Ding:2015fca,Bazavov:2017dus,Bazavov:2018mes,Giordano:2019slo,Bazavov:2020bjn}
around $\mu=\mu_B/3 = 0$, 
simulating at imaginary chemical potential \cite{deForcrand:2002hgr,DElia:2002tig,DElia:2009pdy, 
Cea:2014xva,Bonati:2014kpa,Cea:2015cya,Bonati:2015bha,Bellwied:2015rza,DElia:2016jqh,Gunther:2016vcp,Alba:2017mqu,
Vovchenko:2017xad,Bonati:2018nut,Borsanyi:2018grb,Bellwied:2019pxh,Borsanyi:2020fev},
complex Langevin approach
\cite{Seiler:2012wz,Sexty:2013ica,Aarts:2014bwa,Fodor:2015doa,Sexty:2019vqx,Kogut:2019qmi,Scherzer:2020kiu} and
reweighting \cite{Hasenfratz:1991ax,Barbour:1997ej,Fodor:2001au,Fodor:2001pe,Fodor:2004nz,Giordano:2019gev} 
from $\mu=0$. \footnote{More speculative approaches such as the Lefschetz thimble 
\cite{Cristoforetti:2012su,Cristoforetti:2013wha,Alexandru:2015xva,Alexandru:2016lsn,Nishimura:2017vav} 
and dual variables \cite{Gattringer:2014nxa,Marchis:2017oqi} are currently
not fully developed for lattice QCD.}
All of these approaches share the feature that for infinitesimally small $\mu$ at
fixed spatial volume they are all expected to give correct results. Once $\mu$ is
not infinitesimally small all approaches suffer from uncontrolled systematic uncertainties which 
render them unreliable.

More precisely, the Taylor expansion method for non-infinitesimal $\mu$ requires the computation of high order
$\mu$-derivatives at $\mu=0$. It has the advantage that it provides well-defined physical quantities, namely the cumulants
of the baryon number distribution at $\mu=0$ directly.
However, the measurement itself leads to ever growing cancellations among fermion contractions as the order of the
derivative increases. Furthermore even if a potentially large number of Taylor coefficients are computed with acceptable 
statistical uncertainty, the best case scenario is a reliable estimate of the radius of convergence.
The Taylor expansion method will only provide information within this radius and extrapolation beyond it 
necessarily will involve uncontrolled systematics.

The second extrapolation method mentioned above involves simulating at imaginary $\mu$ where 
there is no sign problem, but the extrapolation from negative $\mu^2$ to positive finite $\mu^2$ 
requires assumptions about the functional form of the $\mu^2$ dependence. This leads again to 
uncontrolled systematic uncertainty in the extrapolation, similar to the case of the Taylor method. 

The third popular method, the complex Langevin approach, is appealing because it is set up at finite $\mu$ directly but the
precise set of necessary and sufficient conditions for it to give the correct result in QCD is so far unknown. 
A set of sufficient conditions for the correctness of the algorithm in general (some a priori, such as the holomorphicity of 
the action, and some a posteriori, such as the quick decay of the field distributions at infinity) has 
been proven \cite{Aarts:2009uq,Aarts:2011ax}, however these conditions are not satisfied in lattice QCD.
Although one may
formulate various tests of incorrectness and the lack of observed such signals may boost confidence in the
correctness of the results, the systematic uncertainties associated with the potential breakdown of the algorithm 
cannot be estimated quantitatively.
Numerical investigations indicate that present
incarnations of the method break down at low temperatures.
Whether an extension of the method capable of simulating also at low temperatures exists is a matter of ongoing research.

Finally the fourth method, reweighting from $\mu=0$, leads to the well-known overlap problem at some finite $\mu$. 
This means that if a suitable weight is found, $w(\mu)$, which may depend on any number of further parameters 
\cite{Fodor:2001au,Fodor:2001pe} beyond
$\mu$, and expectation values are computed via $\langle {\cal O} \rangle_\mu = \langle {\cal O} w(\mu) \rangle_0 /
\langle w(\mu) \rangle_0$, then the histogram of $w(\mu)$ becomes wider and wider for increasing $\mu$. Sampling the tail
of the histogram becomes eventually prohibitively expensive and a reliable error estimate at finite
statistics impossible. Furthermore, there is no sharply defined condition which would signal the presence of the
overlap problem or absence thereof. In practice one may attempt to confirm the lack of the overlap problem from various
statistical observations and may very well obtain reliable results, but the inherent systematic uncertainty will
nevertheless linger.

Our motivation for the present paper is to devise an algorithm which is free of uncontrolled 
systematic uncertainties and has a well-defined
set of conditions for its applicability. In other words we would like to have a trustworthy algorithm in the sense
that results obtained with it are reliable with well-defined statistical uncertainties and have quantifiable, 
controlled systematic uncertainties. We will not solve the sign problem and do not aim to.
Our approach involves the
generation of configurations with the positive fermionic weight $|\re\det D(\mu)|$ where $D(\mu)$ is the Dirac matrix
and the signs ${\rm sign} ( \re \det D(\mu) ) = \pm 1$ are handled by a discrete reweighting.

As an application of the method we perform a study of the conjectured critical end point 
in the $\mu - T$ phase diagram. At $\mu = 0$ QCD has a cross-over thermal transition 
and it is expected that as $\mu$ is increased the transition gets
stronger and eventually at some $\mu = \mu_c$ it becomes a second order phase transition, beyond which at $\mu > \mu_c$
the transition is first order. We would like to unambiguously
observe this strengthening of the transition in a manner which is free of uncontrolled systematic uncertainties. The
present paper will be limited to the unimproved staggered discretization at fixed $N_t = 4$ hence we do not claim to arrive at
continuum results. Nonetheless even at fixed $N_t$ the lattice system, as a well defined statistical physics system, 
may or may not possess a critical end point. This latter question is the one we attempt to address in our paper. 

Note that the mere idea of using $|\re\det D(\mu)|$ as a positive weight to generate configurations
is not new \cite{deForcrand:2002pa, deForcrand:2010ys, Hsu:2010zza}. Actual numerical
simulations with this method were nevertheless only carried out in the canonical approach in the past
\cite{Alexandru:2005ix,Li:2010qf,Li:2011ee}.

The organization of the paper is as follows. In section \ref{pathintegral} we formulate the relevant path integrals in
the presence of a chemical potential and reorganize them in a form which allows for a numerical simulation. We present
our numerical results in section \ref{numericalresults} including our Monte-Carlo algorithm directly at non-zero
$\mu$ as well as the details of our analysis of the leading Fisher (sometimes also called Lee-Yang) 
zeros of the partition function. The volume scaling
of the leading Fisher zeros is used to infer the order of the phase transition at any given non-zero $\mu$. Finally
in section \ref{conclusion} we end with some conclusions and outlook for future work. 

\section{Path integral at finite $\mu$}
\label{pathintegral}

At finite chemical potential the partition function and expectation values are computed as,
\bea
Z(\mu) &=& \int dU \, \det D(U,\mu) e^{-S_g(U)} \nn \\
\langle {\cal O} \rangle_\mu &=& \frac{1}{Z(\mu)} \int dU \, {\cal O}(U)\, \det D(U,\mu) e^{-S_g(U)}\;,
\label{z1}
\eea
where $D(U,\mu)$ is the fermionic Dirac matrix involving all flavors and mass terms and $S_g(U)$ is the gauge action.
As is well-known $\det D(U,\mu)$ is complex for real $\mu \neq 0$, but $Z(\mu)$ is nonetheless real. Hence we may
equivalently write,
\bea
Z(\mu) &=& \int dU \, \re \det D(U,\mu) e^{-S_g(U)}\;. 
\label{z2}
\eea
It is worth emphasizing that taking the real part above is exact and does not introduce any approximation, 
as $Z(\mu)$ in (\ref{z1}) and (\ref{z2}) are exactly identical if charge conjugation invariance holds. For a large
class of observables we may further write,
\bea
\langle {\cal O} \rangle_\mu &=& \frac{1}{Z(\mu)} \int dU\, {\cal O}(U)\, \re \det D(U,\mu) e^{-S_g(U)}\;,
\label{oprop}
\eea
for instance if ${\cal O}(U) = {\cal O}(U^*)$ or if the observable is related to derivatives of $Z(\mu)$ with respect
to a real $\mu$ or mass, etc. In this work we will only be concerned with observables of this type and (\ref{oprop})
will hold. Although the weights are real now the sign
problem of course persists as they can be negative. However one may split the sign $\varepsilon(U,\mu) = \sign\re\det D(U,\mu)$
of the weights from their absolute values and arrive at
\bea
Z(\mu) &=& \int dU \, \varepsilon(U,\mu)\, |\re\det D(U,\mu)| e^{-S_g(U)} \nn \\
\langle {\cal O} \rangle_\mu &=& \frac{1}{Z(\mu)} \int dU \, {\cal O}(U) \, \varepsilon(U,\mu) \, |\re \det D(U,\mu)| e^{-S_g(U)}\;.
\label{epseq}
\eea
Clearly, $|\re\det D(U,\mu)| e^{-S_g(U)}$ is positive and can be used as a weight in importance sampling. 
Configurations will be generated using this weight and the corresponding expectation values will be denoted by
$\langle \ldots \rangle_{abs,\mu}$.
The signs $\varepsilon(U,\mu) = \pm 1$ will be dealt with by a discrete reweighting, 
leading to
\bea
\langle {\cal O} \rangle_\mu &=& \frac{\langle \varepsilon\,{\cal O}
\rangle_{abs,\mu}}{\langle\varepsilon \rangle_{abs,\mu}}\;,
\label{mustuff}
\eea
which is meaningful if the denominator is non-zero. Furthermore, if indeed the denominator is non-zero then the result
is trustworthy as there cannot be any overlap problem, since the only reweighting we need to deal with is a reweighting
with respect to a discrete set; there are only two sectors, those with $\varepsilon(U,\mu)=+1$ and $-1$. The sign
problem is of course still present and it will be signified by the denominator being zero within errors.

To summarize the above, what we have achieved by the formulation (\ref{mustuff}) is that the {\em only} problem is the sign
problem, there is no other uncontrolled systematic which may spoil the result even when the sign problem is not
prohibitively severe. Consequently, we have both a sufficient and necessary condition for the correctness of the
results: if at a given set of parameters and lattice volumes $\langle\varepsilon(U,\mu)\rangle_{abs,\mu}$ is consistent
with zero within statistical uncertainties then we have no result, if on the other hand it is non-zero then whatever the
result is, it is reliable with well-defined statistical errors.

\begin{table}
\begin{center}
\begin{tabular}{|c||c|c|c|c|c|c|c|c|c|c|}
\hline
$a\mu$     & 0.0250 & 0.0500 & 0.0750 & 0.1000 & 0.1250 & 0.1500 & 0.1625 & 0.1750 & 0.1875 & 0.2000 \\
\hline
$\beta_c$ & 5.1870 & 5.1856 & 5.1847 & 5.1827 & 5.1796 & 5.1757 & 5.1739 & 5.1736 & 5.1704 & 5.1686 \\
\hline
\end{tabular}
\end{center}
\caption{The bare couplings used for the 10 different chemical potentials.}
\label{params}
\end{table}

Let us denote the sets of configurations with $\varepsilon(U,\mu)=\pm 1$ by ${\cal U}_\pm(\mu)$, 
which of course depend on $\mu$. Then we have,
\bea
\label{zs}
Z(\mu) &=& Z_+(\mu) - Z_-(\mu) > 0\;, \qquad\quad  Z_\pm(\mu) = \int_{{\cal U}_\pm(\mu)} dU\,|\re\det D(U,\mu)| e^{-S_g(U)} \nn \\
\langle \varepsilon \rangle_{abs,\mu} &=& \frac{Z_+(\mu) - Z_-(\mu)}{Z_+(\mu) + Z_-(\mu)} > 0  \\
\langle {\cal O} \rangle_\mu &=& \frac{ {\cal O}_{+}(\mu) - {\cal O}_{-}(\mu) } {Z_+(\mu) - Z_-(\mu)}\;, \qquad \qquad
{\cal O}_{\pm}(\mu) = \int_{{\cal U}_\pm(\mu)} dU\, {\cal O}(U) |\re\det D(U,\mu)| e^{-S_g(U)}  \nn
\eea
where the inequalities are meant as exact results at infinite statistics while at finite statistics the left
hand sides may of course be consistent with zero within errors.

\section{Numerical results}
\label{numericalresults}

In our simulations we employ the Wilson plaquette gauge action and $2+1$ flavors
of rooted staggered (unimproved) fermions on $N_t = 4$ lattices. 
The spatial lattice sizes are $L/a = 8, 10, 12$ and the fermion masses are set to their physical values 
$a m_{ud} = 0.0092$ and $a m_s = 0.25$. The chemical potential is introduced for the light quarks only, $\mu_u = \mu_d =
\mu$ and $\mu_s = 0$ is set for the strange. The setup is identical to \cite{Fodor:2004nz}.

At each $\mu$ and spatial volume the bare coupling was set to $\beta_c$ as follows. For each $\mu$, 
initial $\beta_{c0}$ values were
taken from \cite{Giordano:2020uvk}. The leading Fisher zeros (see section \ref{fischerzeros}), 
$\beta_1 + i \beta_2$ were measured in shorter 
runs and $\beta_{c0}$ was modified by $\Delta\beta = \beta_1 - \beta_{c0}$ if necessary. Then all 
further production runs were performed at these $\beta_c = \beta_{c0} + \Delta\beta$. The resulting values are
shown in table \ref{params}. From \cite{Giordano:2020uvk} we also glean that the spatial volume dependence of $\beta_c$
is rather mild and in this first exploratory work we set the same bare coupling for all of our 3 spatial volumes.

\begin{figure}
\begin{center}
\includegraphics[width=7.5cm]{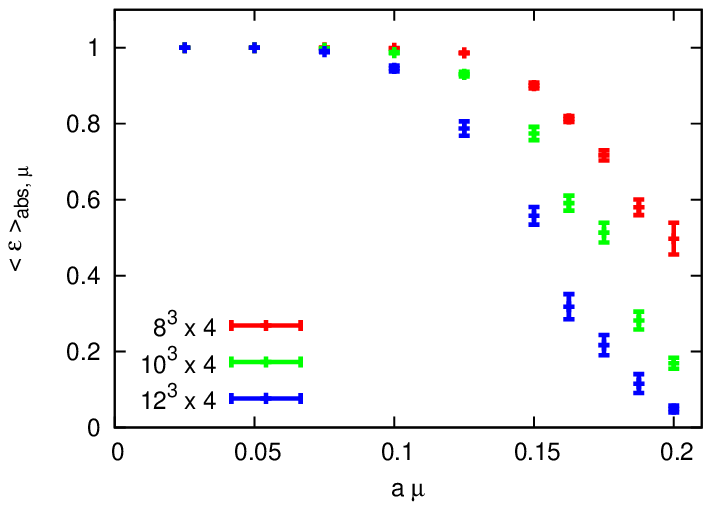} \hspace{-0.5cm} \includegraphics[width=7.5cm]{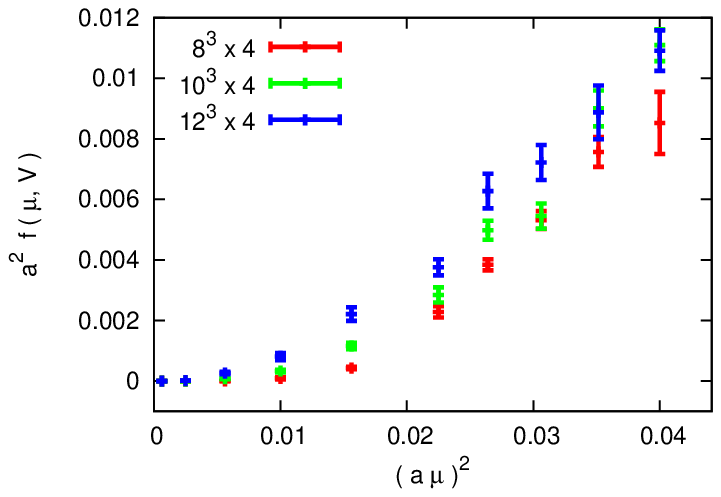}
\end{center}
\caption{Left: the average sign, $\langle \varepsilon \rangle_{abs,\mu}$ for the 3 spatial volumes as a function of the
chemical potential $\mu$. Right: the factor $f(\mu,V)$ parametrizing the average sign; see (\ref{ff}).}
\label{eps}
\end{figure}

\subsection{Monte-Carlo with $\mu > 0$}

We would like to generate configurations with the weight $|\re\det D(U,\mu)| e^{-S_g(U)}$. This is a non-trivial problem
and to our knowledge no pseudo-fermion type construction can be found. What one may still do is rewrite the weight as
\bea
|\re\det D(U,\mu)| e^{-S_g(U)} = \frac{ |\re\det D(U,\mu)|}{ |\re\det D(U,0)|} \det D(U,0) e^{-S_g(U)}\;,
\eea
since $\det D(U,0)$ is real and positive, and utilize a standard (R)HMC algorithm at $\mu=0$ and include the 
$\mu$-dependent ratio in the Metropolis accept/reject step at the end
of the trajectory. This will clearly be an expensive algorithm because the full determinant needs to be computed, but
with the help of the reduced matrix construction the cost is still manageable for the 
lattice volumes we will consider in this paper.

\begin{figure}
\begin{center}
\includegraphics[width=7.5cm]{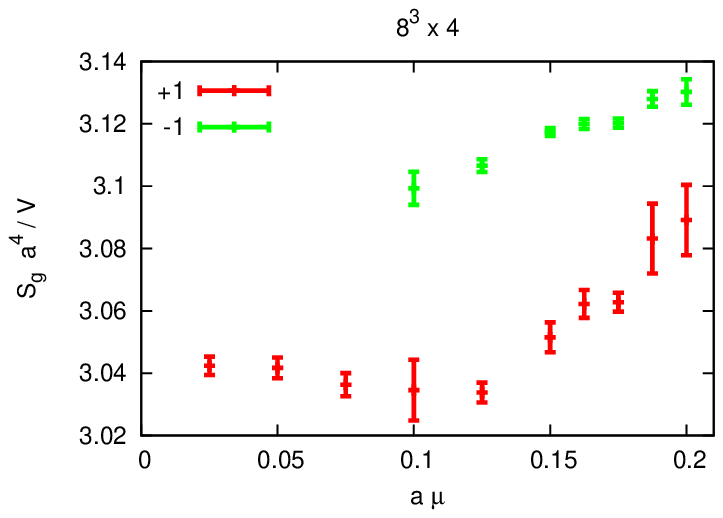} \hspace{-0.5cm} \includegraphics[width=7.5cm]{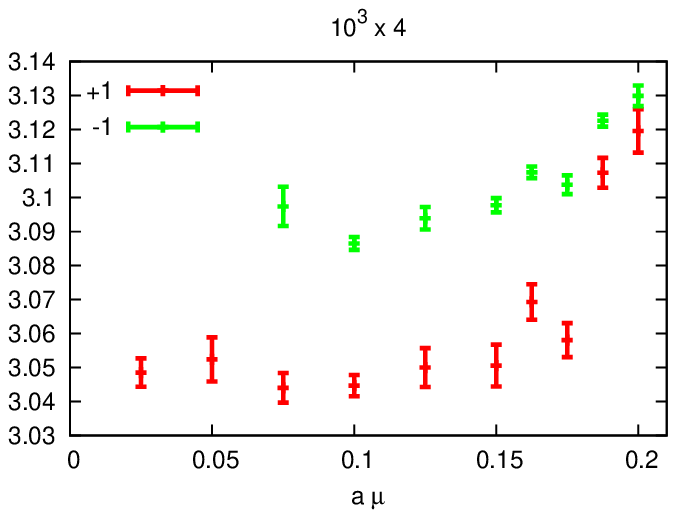} \\
\includegraphics[width=7.5cm]{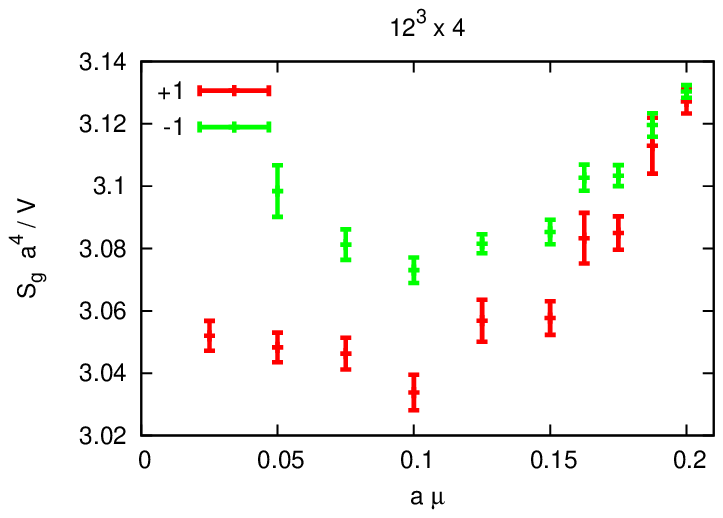}
\end{center}
\caption{The gauge action per unit space time volume $S_g \; a^4 / V = S_g \; (a/L)^3 \; aT $ 
for the 3 spatial volumes and the sectors $+1$ and $-1$ separately. For the low chemical potentials
there are no configurations within the sector $-1$ in our ensembles.}
\label{comp}
\end{figure}

Since we are working with rooted staggered fermions we need to compute the full determinant, its square root, its
real part and then
its sign and absolute value. At finite temperature these steps can most easily be done with the help 
of the reduced matrix \cite{Hasenfratz:1991ax}. This has $6 (L/a)^3$ eigenvalues, $\lambda_i(U)$, and the main utility
of them is that the full staggered determinant can be given at finite chemical potential as,
\bea
\det D_{st}(U,\mu) = \prod_i \left( \lambda_i(U) e^{-\frac{\mu}{2T}} - e^{\frac{\mu}{2T}} \right)\;.
\eea
For the precise definition of the reduced matrix see \cite{Hasenfratz:1991ax}. We will define the square root branch
factor-by-factor in the above product by requiring continuity in $\mu$ or in other words by requiring that in the
$\mu\to0$ limit each factor below goes to unity,
\bea
\frac{ \det D(U,\mu)}{\det D(U,0)} = \left(\frac{\det D_{st}(U,\mu)}{\det D_{st}(U,\mu)}\right)^{1/2} = 
\prod_i  \left( \frac{ \lambda_i(U) e^{-\frac{\mu}{2T}} - e^{\frac{\mu}{2T}} }{ \lambda_i(U) - 1 } \right)^{1/2}\;.
\eea
The branch cut of the square root is placed on the negative real axis.
This procedure fully fixes the complex determinant ratio. The real part, sign and absolute value can then be taken
straightforwardly. Notice that with this procedure the partition function remains real since
$\det D(U^*,\mu) = \det D(U,\mu)^*$ for real $\mu$, and so our approach maintains its validity.

Clearly, if $\mu$ is small the ratio $|\re\det D(U,\mu)| / |\re\det D(U,0)|$ is close to unity and
hence will not affect the Metropolis step much, i.e., a tuned (R)HMC algorithm at $\mu=0$ will perform just as well. On a
given spatial volume as $\mu$ increases the ratio will influence the Metropolis step more and more and will decrease the
acceptance rate. This can be compensated by employing shorter (R)HMC trajectories as this will change the links less and
consequently the change in the ratio with respect to the beginning and end of the trajectory will decrease. In this way
we are able to keep the acceptance rate above $50\%$ for all runs. The shorter trajectories will of course lead to
larger autocorrelation times. Concretely, our estimate of integrated autocorrelation times of our key 
observable (\ref{obs}) are between
50 and 500 depending on $\mu$ and $L/a$. The total number of configurations are between $5 \cdot 10^4$ 
and $2 \cdot 10^5$, leading to a few hundred independent configurations for each simulation point.
We observe that ``tunnelling'' between the $+1$ and $-1$ sectors are frequent,
i.e., the change in the $\mu$-dependent ratio is small enough so that even if the trajectory changes sector we observe
good acceptance.

The crucial measure of whether the results are reliable or not is given by $\langle \varepsilon \rangle_{abs,\mu}$,
i.e., the average sign, which at the same time measures the strength of the sign problem itself. Since 
$\langle \varepsilon \rangle_{abs,\mu} \to 1$ as $\mu \to 0$ we parametrize it as,
\bea
\label{ff}
\langle \varepsilon \rangle_{abs,\mu} = e^{-V \mu^2 f(\mu,V)}
\eea
with the 4-volume $V = L^3/T$ and show in figure \ref{eps}
both $\langle \varepsilon \rangle_{abs,\mu}$ as well as $f(\mu,V)$. Clearly, $f(\mu,V)$ depends
mildly on $V$ but does depend non-trivially on $\mu$. As can be seen the volumes $L/a=8,10,12$ and chemical potentials 
$a\mu \leq 0.2$ are safely in the region where  $\langle \varepsilon \rangle_{abs,\mu}$ is several standard deviations away
from zero, hence the sign reweighting (\ref{mustuff}) can be performed without issues. In particular, as emphasized,
there is no overlap problem to contend with.

It is worth exploring what the effect of the sign reweighting is on some observables, more precisely how different
some observables are in the $+1$ and $-1$ sectors. As an example we show the
gauge action per unit space time volume in figure \ref{comp} as a function of the chemical potential separately for the
two sectors.

\subsection{Fisher zeros}
\label{fischerzeros}

\begin{figure}
\begin{center}
\includegraphics[width=7.5cm]{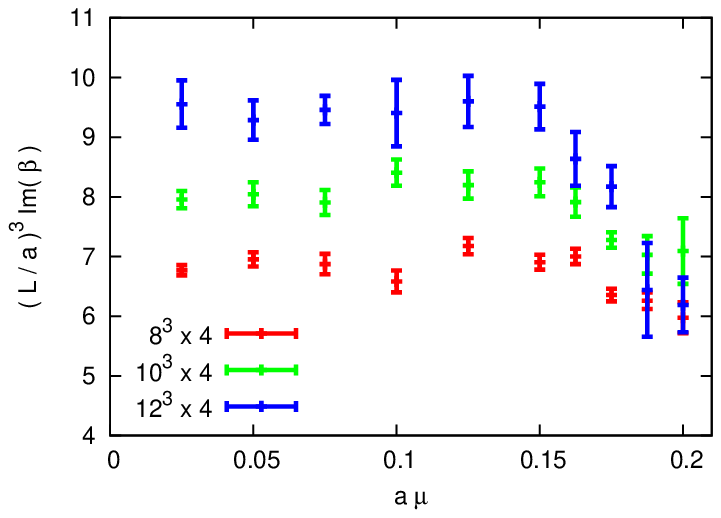} \hspace{-0.5cm} \includegraphics[width=7.5cm]{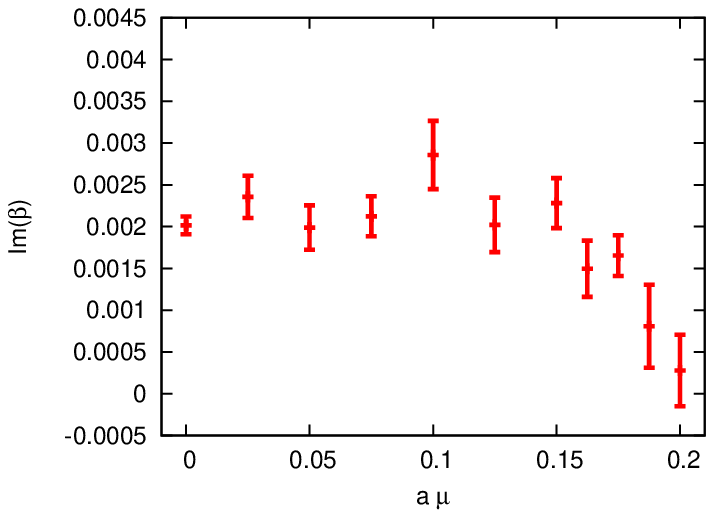}
\end{center}
\caption{Left: the measured imaginary parts of the Fisher zeros scaled by the spatial volume. Right: 
the infinite volume extrapolated imaginary part of the Fisher zeros. The result at $\mu = 0$ was obtained
using the standard (R)HMC algorithm.}
\label{inf}
\end{figure}

Once it has been determined which volumes and chemical potentials allow for the application of the sign reweighting
(\ref{mustuff}) we are able to compute observables. Since our primary interest is the order of the phase transition as a
function of $\mu$ we will compute the Fisher zeros in the bare coupling $\beta$, i.e.,
we will look for complex bare couplings such that $Z(\mu,\beta) = 0$ at given $\mu$ and volume; see (\ref{zs}). 
This amounts to measuring the observables
\bea
{\cal O}(U) = e^{-(\beta-\beta_c) \frac{S_g(U)}{\beta_c}}
\label{obs}
\eea
for complex $\beta$, assuming the simulation was done at (real) bare coupling $\beta_c$. Since ${\cal O}(U^*) = {\cal
O}(U)$ our method can be applied without problems. More precisely, since $Z(\mu,\beta)$ has several zeros
as a function of complex $\beta$, we will be looking for the one closest to the real axis, which in every run happens to
coincide with the one closest to $(\beta_c,0)$ in the complex plane as well. This zero will be called the leading zero.

\begin{figure}
\begin{center}
\includegraphics[width=5.8cm]{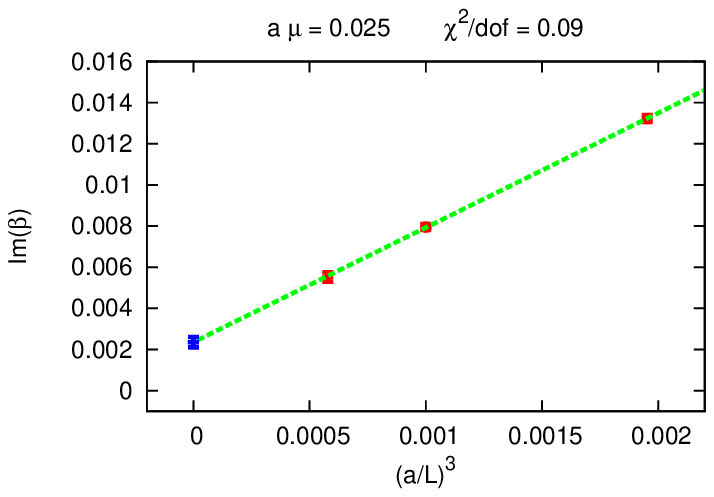}  \includegraphics[width=5.8cm]{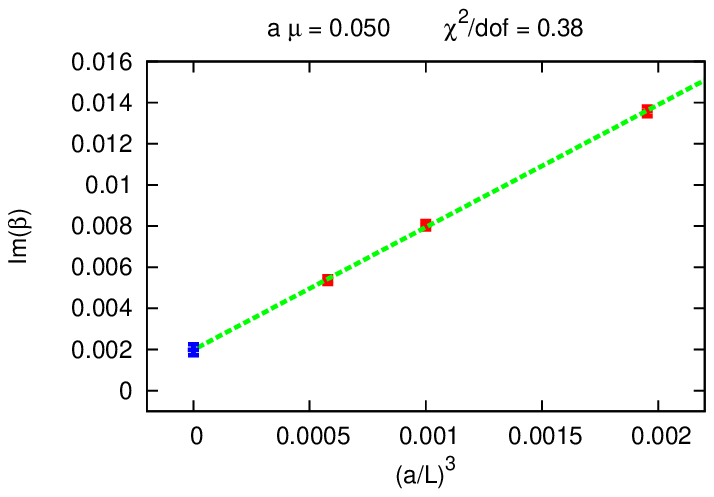} \\
\includegraphics[width=5.8cm]{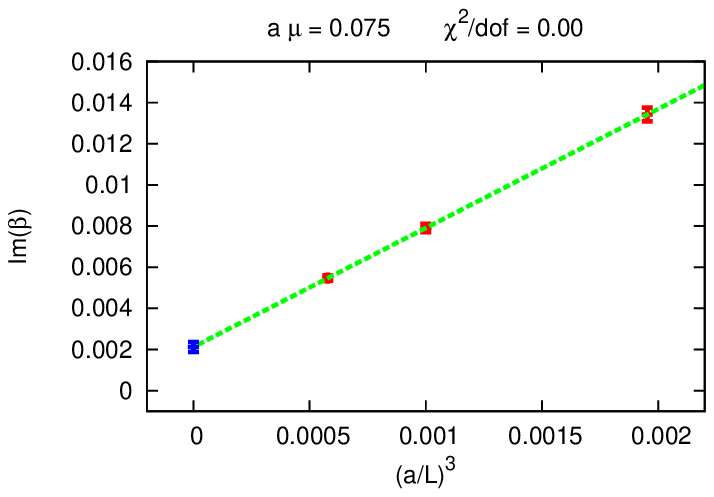}  \includegraphics[width=5.8cm]{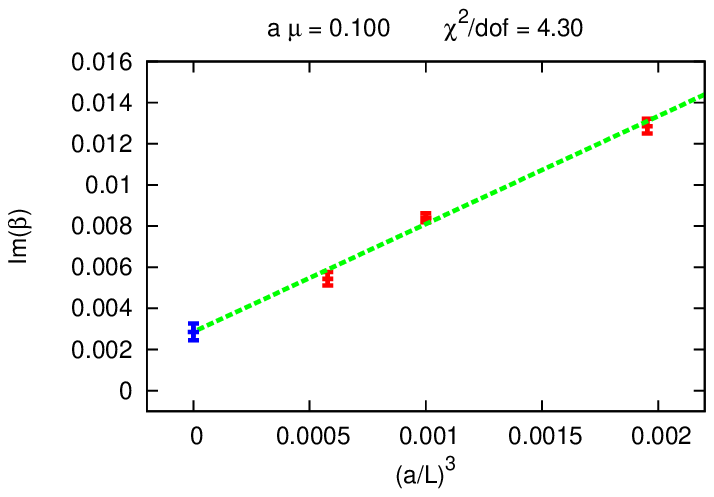} \\
\includegraphics[width=5.8cm]{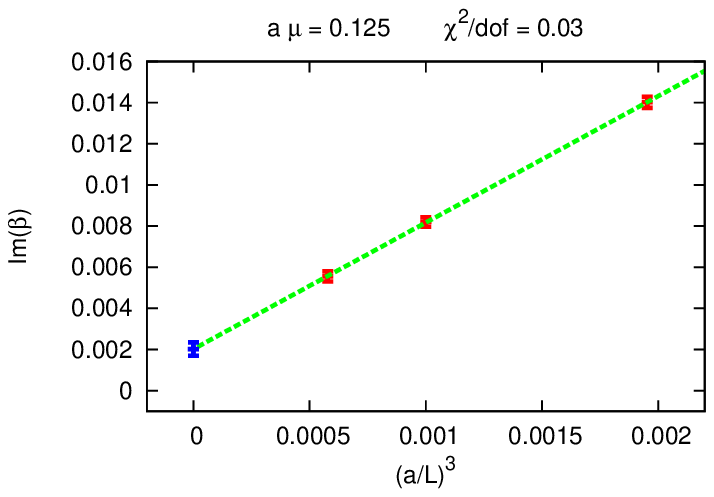}  \includegraphics[width=5.8cm]{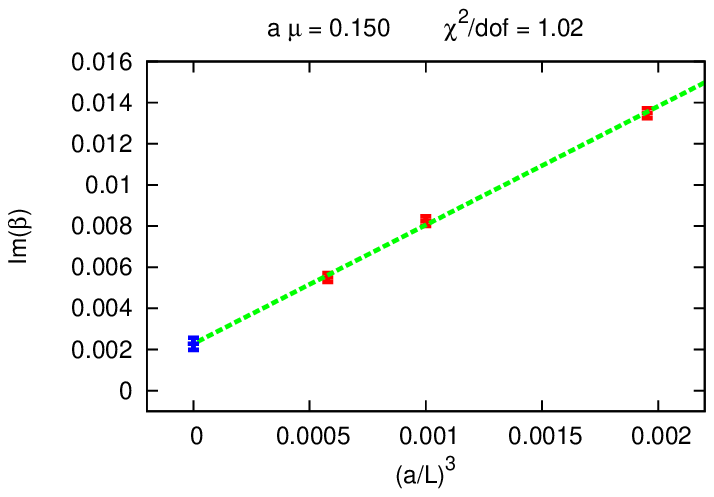} \\
\includegraphics[width=5.8cm]{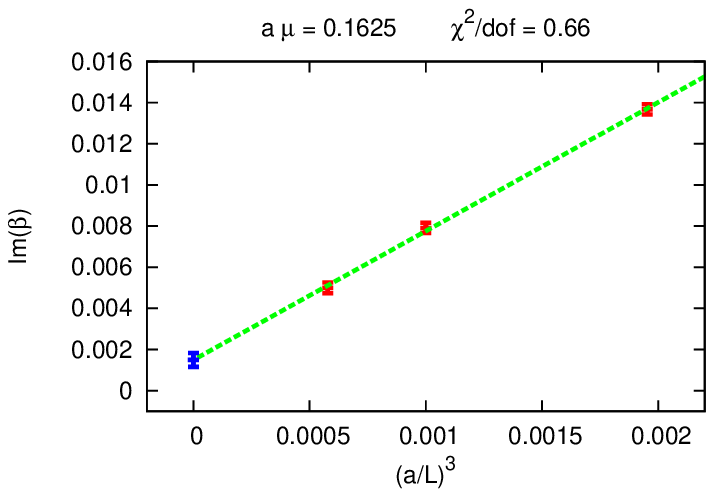} \includegraphics[width=5.8cm]{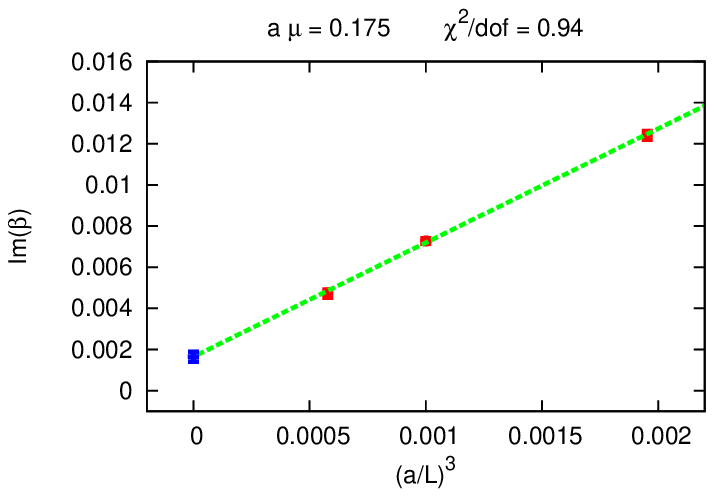} \\
\includegraphics[width=5.8cm]{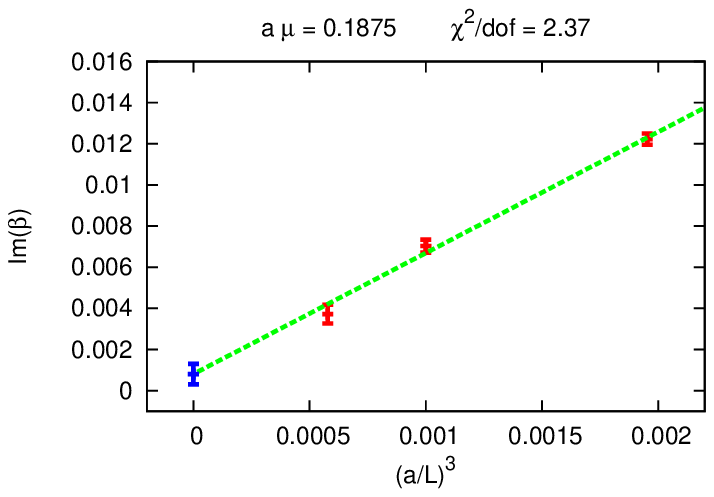} \includegraphics[width=5.8cm]{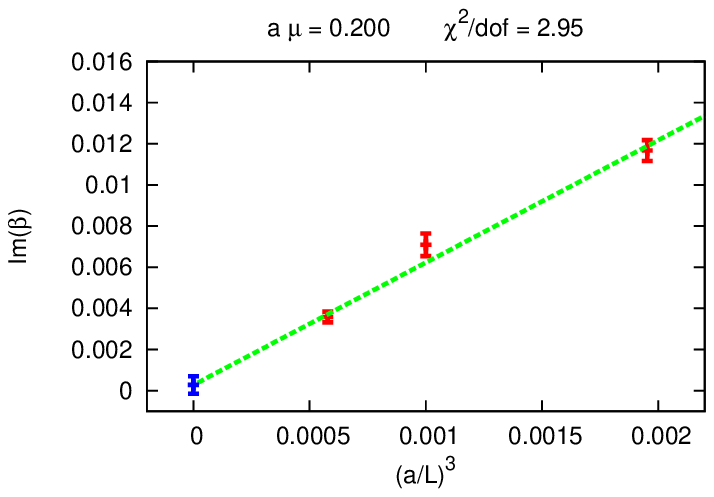}
\end{center}
\caption{The infinite volume extrapolations (\ref{exp}) of the 
imaginary parts of the leading Fisher zeros at the various chemical
potentials we have simulated at.}
\label{infall}
\end{figure}

The volume scaling of $\im \beta$ determines the order of the transition: if
$\im\beta \to const$ as $L/a \to \infty$ the transition is a cross-over, if $\im\beta \sim a^3/L^3$ the transition is
first order and finally if $\im\beta \sim (a/L)^\alpha$ with a non-trivial exponent $\alpha > 0$ the transition is second
order. Although these leading order expressions are unambiguous in all three cases, the subleading terms are a priori not
known. Since we know that at $\mu=0$ the transition is a cross-over for physical quark masses, it is generally expected
that for small $\mu$ it will stay a cross-over. At fixed $\mu > 0$ the imaginary part of the leading Fisher zero is
then extrapolated to infinite volume via,
\bea
\im\beta &=& A + B \, ( a /L )^3\;,
\label{exp}
\eea
where the exponent $3$ in the subleading term is merely an ansatz. 
In this first study we only simulated at 3 volumes, $L/a = 8,10,12$ and hence we are unable to fit the exponent 
of the subleading term simultaneously with $A$ and $B$. Empirically, we do find that the above fit function
provides acceptable statistical fits for our choice of chemical potentials.

The existence of a critical end point would suggest that
$\im\beta_\infty(\mu) = A(\mu)$ is a decreasing function of $\mu$ and as $\mu\to\mu_c$ we have
$A(\mu) \to 0$.

The real part of the leading Fisher zero on the other hand may be used to define the critical coupling. The simulations
we performed at particular values of $\beta_c = \beta_c(\mu)$ and we have checked that for $a\mu > 0.1$ the differences
$\Delta\beta = \re \beta - \beta_c$ are deviating from zero less than $3\sigma$ and rarely beyond $1.5\sigma$.
Note that a smooth cross-over means that different observables may lead to different definitions of the pseudo-critical
coupling. 

The measured imaginary parts of the Fisher zeros are shown in the left panel of figure \ref{inf}. 
The extrapolations to infinite volume using $L/a=8,10,12$ are shown in figure \ref{infall} together with
the resulting $\chi^2/dof$ values of the fits. Out of the 10 extrapolations the largest $\chi^2/dof$ values are at
$a\mu=0.1,\; 0.1875,\; 0.2$ and are $4.3,\; 2.37,\; 2.95$. Note that $dof=1$ and even the largest $\chi^2/dof=4.3$ leads
to a q-value of $4\%$.
The resulting $\im(\beta_\infty(\mu))$ as a function of $\mu$ is finally shown in the right panel of figure \ref{inf}.

The most important result from our investigation can be gleaned from figure \ref{inf}. Both at finite volumes and
correspondingly in infinite volume the imaginary part of the relevant Fisher zero is decreasing as the chemical
potential becomes sufficiently large. More precisely, the infinite volume extrapolated result shows that the imaginary part of the
leading Fisher zero is more or less flat up to $a\mu \sim 0.15$ and a sharp decrease is observed for $0.15 \leq a\mu
\leq 0.2$. The observed flatness agrees within errors with the slight increase seen in \cite{Fodor:2004nz,Giordano:2020uvk}, 
and cannot be significantly distinguished from it with the currently available statistics.
This means that in the range of chemical
potentials where our results are reliable with trustworthy statistical errors, i.e., $\langle \varepsilon
\rangle_{abs,\mu} \neq 0$, we are able to conclude with high statistical significance that the leading singularity of
$\log Z$ is eventually moving closer and closer to the real axis. In fact, the location of the singularity is consistent with a
real value at $a\mu \sim 0.2$.
This in turn means that the strength of the transition is
eventually increasing and very suggestive that a true phase transition occurs at around $a \mu \sim 0.2$. This corresponds to
$\mu / T_c \sim 0.8$, in agreement with \cite{Fodor:2004nz}, however the latter result 
should be interpreted with caution since, as we explained, we do not expect
the fixed $N_t = 4$ results to be particularly close to the continuum with our chosen discretization. Nonetheless our
results are trustworthy in the well defined statistical model given by the $N_t = 4$ lattice system.

\section{Conclusion and outlook}
\label{conclusion}

In this paper we have introduced a new technique for evaluating the path integral at finite baryon chemical potential.
The approach involves generating configurations by the absolute value of the real part of the fermionic determinant and
taking the signs into account by a discrete reweighting. The first step necessitates the evaluation of the full
determinant during the Monte-Carlo simulation which makes the algorithm rather costly but still manageable for $8^3
\times 4$, $10^3 \times 4$ and $12^3 \times 4$ which are the volumes we used. The second step, the discrete reweighting
by the sign of the real part of the fermionic determinant, is a fully controlled step provided the average sign
is several standard deviations away from zero, i.e., the sign problem is not too severe. 
This requirement can be easily monitored and once it is fulfilled, the results are completely trustworthy with
well-defined statistical errors. This feature is the main advantage of our method. It improves on traditional
reweighting in $\mu$ and/or some other continuous parameter because in that case the notorious overlap problem may
invalidate the results even though a naive application of the reweighting formula $\langle {\cal O}\rangle_\mu = \langle
{\cal O} w \rangle_0 / \langle w \rangle_0$ seemingly presents no problems.

Since our main interest was the order of the thermal phase transition as a function of the chemical potential, we have
determined the first few Fisher zeros and the volume scaling of the leading one (the one closest to the real axis)
for 10 choices of $\mu$ in the range $0.025 \leq a \mu \leq
0.2$. We have observed that the strength of the phase transition stays flat within errors for $0 < a\mu < 0.15$ and increases
sharply for $0.15 < a\mu < 0.2$, signified by the decrease in the imaginary part of the leading Fisher zero. The
infinite volume extrapolation of the leading Fisher zero at $a\mu \sim 0.2$, corresponding to $\mu_B / T \sim 2.4$, is in 
fact consistent with a true phase transition, i.e., the imaginary part is consistent with zero.

There are however several avenues to improve on our work in the future. First, we have performed simulations at fixed
$N_t = 4$, i.e., we have not addressed the continuum limit at all; simulations with larger temporal extents would be
necessary in order to do so. Once the continuum behavior is investigated it might be worthwhile to use an improved
action, both for the gauge and fermionic actions. In the present work we have used the Wilson plaquette gauge action and
unimproved staggered fermions. The motivation was to replicate the setup of \cite{Fodor:2004nz} where the critical end
point was investigated using traditional reweighting in $\mu$. It is worth noting that even though the unimproved
staggered discretization on $N_t = 4$ lattices is far from the continuum, it is a well-defined lattice statistical physics model
with a sign problem. Hence it makes perfect sense to study it in order to gain valuable insight into the sign problem
in general.

Second, the volume scaling of the Fisher zeros is of central importance and our ansatz (\ref{exp}) was simply
motivated by empirically observing good statistical fits as well as the fact that we only had data on 3 volumes.
Hence we were unable to fit all 3 parameters, $A,B$ and $C$ in the general form,
\bea
\im\beta = A + B \, ( a / L )^C\;,
\eea
which would otherwise be the justified procedure. Once an additional volume $14^3 \times 4$ is added, the exponent $C$
could be determined or at least constrained. 

Third, we have set the quark masses to their physical values at $\beta = \beta_c$ at $\mu = 0$ and have not changed them
for $\mu > 0$ along the line of constant physics. Even though the effect is expected to be negligible relative to
other sources of errors, in future work we do plan to follow the line of constant physics for $\mu > 0$.

Fourth, we have included the chemical potential at the quark level as $\mu_u = \mu_d = \mu_B / 3 = \mu$ 
and $\mu_s = 0$ which corresponds to $\mu_S = \mu_B / 3$. Nonetheless our method can be
trivially modified to include other chemical potential assignments, e.g., strangeness neutrality $\langle S \rangle = 0$
or $\mu_S = 0$.

Finally we mention that the recently introduced geometric matching procedure \cite{Giordano:2019gev} provides a
new rooting procedure at finite $N_t$ which is nonetheless expected to agree with the one followed in this paper towards
the continuum limit. We repeated the determination of the leading Fisher zeros using geometric matching and found that
they agree with the ones presented in this paper within statistical uncertainties. This type of cross-check will be
especially useful for future studies targeting the continuum limit.

\section*{Acknowledgments}
This work was partially supported by the Hungarian National
Research, Development and Innovation Office - NKFIH grant KKP-126769.
AP is supported by
the J\'anos Bolyai Research Scholarship of the Hungarian Academy of
Sciences and by the \'UNKP-19-4 New National Excellence Program of
the Ministry for Innovation and Technology.

\end{document}